\documentclass[%
 reprint,
 superscriptaddress,
 nofootinbib,
 twocolumn, 
 amsmath,amssymb,
 aps,
prb,
]{revtex4-2}

\usepackage[colorlinks = true,
            linkcolor = blue,
            urlcolor  = red,
            citecolor = red,
            anchorcolor = blue]{hyperref}
\usepackage[T1]{fontenc}
\usepackage[utf8]{inputenc}
\usepackage{graphicx}
\usepackage{dcolumn}
\usepackage{color}
\usepackage{amsmath}
\usepackage{appendix}
\usepackage{MnSymbol}

\usepackage{glossaries}
\newacronym{CDW}{CDW}{charge-density-wave}
\newacronym{BOW}{BOW}{bond-order wave}
\newacronym{DQMC}{DQMC}{determinant quantum Monte Carlo}
\newacronym{QMC}{QMC}{quantum Monte Carlo}
\newacronym{SSH}{SSH}{Su-Schrieffer-Heeger}
\newacronym{2D}{2D}{two-dimensional}
\newacronym{FS}{FS}{Fermi surface}

\begin{document}

\preprint{}
\title{A comparative study of the superconductivity in the Holstein and optical Su–Schrieffer–Heeger models}

\author{Andy~Tanjaroon~Ly}
\affiliation{Department of Physics and Astronomy, The University of Tennessee, Knoxville, TN 37996, USA}
\affiliation{Institute of Advanced Materials and Manufacturing, The University of Tennessee, Knoxville, TN 37996, USA\looseness=-1} 

\author{Benjamin~Cohen-Stead}
\affiliation{Department of Physics and Astronomy, The University of Tennessee, Knoxville, TN 37996, USA}
\affiliation{Institute of Advanced Materials and Manufacturing, The University of Tennessee, Knoxville, TN 37996, USA\looseness=-1} 

\author{Sohan~Malkaruge~Costa}
\affiliation{Department of Physics and Astronomy, The University of Tennessee, Knoxville, TN 37996, USA}
\affiliation{Institute of Advanced Materials and Manufacturing, The University of Tennessee, Knoxville, TN 37996, USA\looseness=-1} 

\author{Steven~Johnston}
\affiliation{Department of Physics and Astronomy, The University of Tennessee, Knoxville, TN 37996, USA}
\affiliation{Institute of Advanced Materials and Manufacturing, The University of Tennessee, Knoxville, TN 37996, USA\looseness=-1}

\date{\today}

\begin{abstract}
Theoretical studies suggest that Su-Schrieffer-Heeger-like electron-phonon ($e$-ph) interactions can mediate high-temperature bipolaronic superconductivity that is robust against repulsive electron-electron interactions. Here we present a comparative analysis of the pairing and competing charge/bond correlations in the two-dimensional Holstein and optical Su-Schrieffer-Heeger (SSH) models using numerically exact determinant quantum Monte Carlo. We find that the SSH interactions support light bipolarons and strong superconducting correlations out to relatively large values of the $e$-ph coupling $\lambda$ and densities near half-filling, while the Holstein interaction does not due to the formation of heavy bipolarons and competing charge-density-wave order. We further find that the Holstein and SSH models have comparable pairing correlations in the weak coupling limit for carrier concentrations $\langle n \rangle \ll 1$, where competing orders and polaronic effects are absent. These results support the proposal that SSH (bi)polarons can support superconductivity to larger values of $\lambda$ in comparison to the Holstein polaron, but that the resulting $T_\mathrm{c}$ gains are small in the weak coupling limit. We also find that the SSH model's pairing correlations are suppressed after including a weak on-site Hubbard repulsion. These results have important implications for identifying and engineering bipolaronic superconductivity.
\end{abstract}

\maketitle

\section{Introduction}

Identifying the highest superconducting transition temperature $T_\mathrm{c}$ that can be realized for a given pairing interaction persists as a significant unsolved problem~\cite{Cohen1972comments, Leavens1975least, Allen1975transition, Carbotte1990properties, Paiva2004critical, Moussa2006two, Berg2008route, Varma2012considerations, Lin2016optimal, Esterlis2018bound, Hazra2019bounds}. In particular, there is a long-standing discussion over whether the $T_\mathrm{c}$ arising from the electron-phonon ($e$-ph) interaction is bounded~\cite{Cohen1972comments, Leavens1975least, Allen1975transition, Carbotte1990properties, Esterlis2018bound}. While Eliashberg theory predicts that $T_\mathrm{c}$ increases indefinitely with the dimensionless $e$-ph coupling $\lambda$~\cite{Allen1975transition}, lattice instabilities or competing phases like \gls*{CDW} order are expected to ultimately cut off this growth ~\cite{Leavens1975least, Moussa2006two}. For example, even if one could avoid a lattice instability, the formation of heavy (bi)polarons should suppress superconductivity in the strong coupling limit~\cite{Alexandrov2001breakdown}. These expectations have been recently confirmed by numerical calculations for the Holstein model~\cite{Lin2016optimal, Esterlis2018bound, Esterlis2018breakdown}, suggesting polaron formation sets an intrinsic limit on phonon-mediated superconductivity. 

In this context, \gls*{SSH}-like ($e$-ph) interactions, where the atomic motion modulates the electronic hopping integrals~\cite{Barisic1970tightbinding, Su1979solitons}, have attracted substantial interest~\cite{Sous2018light, Li2020quantum,  Yam2020peierls, Xing2021quantum, Nocera2021bipolaron, Feng2022phase, sous2022bipolaronic, Zhang2023bipolaronic, CohenStead2023hybrid, banerjee2023ground, goetz2023phases}. Recent theoretical studies in the dilute limit suggest that this interaction can produce strongly bound yet light bipolarons less prone to localization~\cite{Sous2018light}, in contrast to the Holstein model. It has also been proposed ~\cite{Zhang2023bipolaronic} that a dilute gas of such bipolarons will have an instability toward a superconducting state with $T_\mathrm{c}$'s much larger than those estimated for the Holstein model~\cite{Esterlis2018bound}. Importantly, pairing mediated by the \gls*{SSH} interaction is also believed to be robust against the inclusion of repulsive $e$-$e$ interactions~\cite{Zhang2023bipolaronic, sous2022bipolaronic}, thus providing a means to bypass the proposed limits of conventional superconductivity. 

The $T_\mathrm{c}$ estimates obtained in Ref.~\cite{Zhang2023bipolaronic} are based on simulations of the \gls*{2D}  bond-\gls*{SSH} model, where the properties of an isolated bipolaron on a lattice are computed with numerically exact \gls*{QMC} methods~\cite{Zhang2022bond}. The authors then argue that the system in the dilute limit can be regarded as a gas of bipolarons and mapped onto the problem of a gas of hard-core bosons, where $T_\mathrm{c}$ was identified with the superfluid transition temperature~\cite{Pilati2008critical}. This approach should be accurate in the dilute limit, where the bipolarons do not overlap significantly~\cite{Zhang2023bipolaronic}. 
While these results suggest that \gls*{SSH} interactions may be able to mediate high-$T_\mathrm{c}$ superconductivity, many materials hosting these types of interactions are far from the dilute limit. As the carrier concentration increases, one naturally expects other ordering tendencies to appear that will compete with superconductivity, e.g., a \gls*{CDW} phase for the Holstein model~\cite{Scalettar1989competition, Marsiglio1990pairing, Noack1991charge, Hohenadler2004quantum, Bradley2021superconductivity,  Dee2023charge, CohenStead2020langevin}. For the \gls*{2D} bond SSH model, studies have identified insulating \gls*{CDW} order~\cite{Li2020quantum, CohenStead2023hybrid}, a \gls*{BOW}~\cite{Xing2021quantum} phase, and even weak antiferromagnetism~\cite{Feng2022phase, Cai2022robustness} near half-filling. It is thus desirable to assess the strength of the relevant pairing correlations in the SSH model and any competing instabilities across a wide range of carrier concentrations and model parameters. 

Here we present a comparative study of the \gls*{2D} Holstein~\cite{Holstein1959studies}, bond~\cite{Sengupta2003Peierls}, and optical~\cite{Capone1997small} SSH models using numerically exact and sign-problem free \gls*{DQMC} simulations. Focusing on the weak coupling limit, where the linear approximation for the \gls*{SSH} interaction is valid~\cite{li2022suppressed, banerjee2023ground}, we find that the models have strong \gls*{CDW} (Holstein) or \gls*{BOW} (bond-/optical-SSH) correlations near half-filling. These phases are suppressed upon doping and give way to strong superconducting correlations in the antiadiabatic limit, where the phonon energy is much larger than the electronic hopping ($\Omega/t \gg 1$, $\hbar = 1$). In this case, we further find that the Holstein and optical-SSH models produce comparable pairing correlations over a wide range of carrier concentrations. In contrast, correlations in the bond-SSH model are notably weaker. For smaller phonon energies $(\Omega/t = 0.5)$, we find that the \gls*{SSH} models can support robust pairing in proximity to their competing bond order at half-filling, while the Holstein model is prone to the formation of heavy bipolarons in the same window. Further doping into the dilute limit shows that all three models have weak but comparable superconducting correlations. Finally, we find that including a modest Hubbard $U=t/2$ suppresses the pairing correlations of the doped SSH models. 

These results provide new insight into the search for bipolaronic superconductors and suggest that one should look toward weakly correlated materials with carrier concentrations close to their relevant competing order. 

\section{Model \& Methods}\label{sec:methods}
We study the single-band Holstein and bond-/optical-SSH models defined on a \gls*{2D} square lattice. The Hamiltonian for each of these models is partitioned as $\hat{H} = \hat{H}_{e} + \hat{H}_{\text{ph}} + \hat{H}_{e\text{-ph}}$, where $\hat{H}_{e}$ describes the electronic subsystem, $\hat{H}_{\text{ph}}$ describes the lattice subsystem, and 
$\hat{H}_{e\text{-ph}}$ describes the coupling between the two. For all three models, we take 
\begin{equation}
    \hat{H}_e =
    -t\sum_{{\bf i}, \nu,\sigma} (\hat{c}^\dagger_{{\bf i}+{\bf a}_\nu,\sigma}\hat{c}^{\phantom\dagger}_{{\bf i},\sigma} + \textrm{h.c.})
    - \mu\sum_{{\bf i},\sigma} \hat{n}_{{\bf i},\sigma}, 
\end{equation}
$\hat{c}^\dagger_{{\bf i},\sigma}$ ($\hat{c}^{\phantom\dagger}_{{\bf i},\sigma}$) creates (annihilates) a spin-$\sigma$ ($=\uparrow,\downarrow$) electron at lattice site ${\bf i}$,  $\hat{n}_{\mathbf{i},\sigma} = \hat{c}^\dagger_{\mathbf{i},\sigma}\hat{c}^{\phantom\dagger}_{\mathbf{i},\sigma}$ is the spin-$\sigma$ electron number operator for site $\mathbf{i}$, and $\text{h.c.}$ denotes the Hermitian conjugate. The sum over $\nu$ runs over the $\hat{x}$ and $\hat{y}$ spatial dimensions, with ${\bf a}_x=(a,0)$ and ${\bf a}_y = (0,a)$. Lastly, $t$ is the nearest neighbor hopping integral, and $\mu$ is the chemical potential.

The Holstein model describes the phonons using local harmonic oscillators  
\begin{equation}
\hat{H}_{\text{ph}} = \sum_{\bf i} \left(\frac{1}{2M_{\rm h}}\hat{P}_{\bf i}^2 + \frac{1}{2}M_{\rm h}^{\phantom 2}\Omega_{\rm h}^2\hat{X}_{\bf i}^2\right), 
\end{equation}
where $M_{\rm h}$ is the ion mass, $\Omega_{\rm h}$ is the phonon energy, and $\hat{X}_{i}$ ($\hat{P}_i$) is the position (momentum) operator for the atom at site ${\bf i}$. For the bond- and optical-SSH models, we introduce harmonic oscillators describing motion along each of the $\nu$ directions of the lattice such that 
\begin{equation}
\hat{H}_\mathrm{ph}= \sum_{\mathbf{i},\nu}\bigg( \frac{\hat{P}_{\mathbf{i},\nu}^2}{2M_{\rm (b,o)}}+\frac{1}{2} M_{\rm (b,o)}\Omega^2\hat{X}_{\mathbf{i},\nu}^2 \bigg), 
\end{equation}
where $M_{\rm (b,o)}$ is the (effective) ion mass in the bond- and optical-SSH models, respectively. In the optical-SSH model, these oscillators describe the displacement of the atoms  themselves~\cite{Capone1997small}. In the bond model, they describe the change in the \textit{relative} distance between the atoms at sites ${\bf i}$ and ${\bf i}+ {\mathbf a}_{\nu}$~\cite{Weber2015excitation}. 

The coupling between the two sub-systems for the Holstein model is given by the usual local interaction $\hat{H}_{e\text{-ph}} = \alpha_\text{h}\sum_{{\bf i},\sigma}\hat{X}_{\bf i}(\hat{n}_{{\bf i},\sigma}-\tfrac{1}{2})$, where $\alpha_\mathrm{h}$ parameterizes the strength of the $e$-ph coupling. Conversely, for the SSH models, the interaction modulates the nearest neighbor hopping integrals, with 
\begin{equation}
    \begin{aligned}
    \hat{H}_{e\text{-ph}} &= \alpha_{\rm b} \sum_{\mathbf{i},\nu,\sigma} \hat{X}_{\mathbf{i},\nu} (\hat{c}^\dagger_{{\bf i}+{\bf a}_\nu,\sigma}\hat{c}^{\phantom\dagger}_{{\bf i},\sigma} + \textrm{h.c.})~\text{and}\\
    \hat{H}_{e\text{-ph}}&= \alpha_{\textrm{o}}\sum_{\mathbf{i},\nu,\sigma}(\hat{X}_{\mathbf{i}+\mathbf{a}_\nu,\nu}-\hat{X}_{\mathbf{i},\nu}) (\hat{c}^\dagger_{{\bf i}+{\bf a}_\nu,\sigma}\hat{c}^{\phantom\dagger}_{{\bf i},\sigma} + \textrm{h.c.})
    \end{aligned}
\end{equation}
for the bond- and optical-SSH models, respectively. In the Holstein model, the electrons couple to a single dispersionless optical phonon branch via a momentum independent $e$-ph coupling constant. In contrast, they couple to two dispersionless optical phonon branches via momentum-dependent $e$-ph coupling constants in the optical- and bond-SSH models. This difference has important implications for properly equating each model's dimensionless $e$-ph coupling constant $\lambda$, as discussed in Ref.~\cite{Sohan} and Appendix~\ref{sec:lambda}. 

We solve all three models on $N = L\times L$ site square lattices with periodic boundary conditions using \gls*{DQMC}~\cite{White1989numerical, CohenStead2022fast}. Throughout, we set $t = M = M_\mathrm{o} = M_\mathrm{b} = a = 1$, and choose the $e$-ph coupling constants $\alpha$ such that the momentum-averaged dimensionless coupling $\lambda$ is the same for all three models (see App.~\ref{sec:lambda}). Finally, we note that the sign of the effective hopping integrals can change in the SSH models when the lattice displacements are large enough~\cite{Nocera2021bipolaron, banerjee2023ground}; our implementation does not reject moves producing such configurations, as described in Appendix~\ref{sec:sign_change}. 

We assess the models' ordering tendencies by measuring the relevant susceptibilities 
\begin{equation}
    \chi_{\gamma}(\mathbf{q})=\frac{1}{N}\int_{0}^{\beta}\sum_{{\bf i},{\bf j}}e^{-\mathrm{i}\mathbf{q}\cdot(\mathbf{R}_{\bf i}-\mathbf{R}_{\bf j})}\left\langle \hat{O}^{\phantom\dagger}_{\gamma,{\bf i}}(\tau)\hat{O}_{\gamma,{\bf j}}^{\dagger}(0)\right\rangle \mathrm{d}\tau,  
\end{equation}
where $\hat{O}_{\gamma,{\bf i}}$ is a local operator. For the superconducting correlations, we set $\hat{O}_{\text{p},\mathbf{i}}  = \hat{c}_{{\bf i},\uparrow}\hat{c}_{{\bf i},\downarrow}$ to measure local $s$-wave pairing and $\hat{O}_{\text{p},\mathbf{i}}^{s^*}  = \frac{1}{2}\sum_{\nu}(\hat{c}_{{\bf i},\uparrow}\hat{c}_{{\bf i+a_{\nu}},\downarrow} + \hat{c}_{{\bf i},\uparrow}\hat{c}_{{\bf i-a_{\nu}},\downarrow})$ to measure extended $s$-wave pairing. For the charge and bond-order correlations, we take $\hat{O}_{\text{c},\mathbf{i}} = \sum_\sigma \hat{n}_{{\bf i},\sigma}$ and $\hat{O}_{\mathrm{b},\mathbf{i}}  =  \sum_{\sigma}(\hat{c}^\dagger_{{\bf i},\sigma}\hat{c}^{\phantom\dagger}_{{\bf i}+{\bf a}_\nu,\sigma} + \textrm{h.c.})$, respectively. 

To assess the polaronic tendencies in each model, we also measured the carrier's effective mass at the Fermi level 
\begin{equation}\label{eq:mstar}
    \frac{m^*({\bf k})}{m} = 1 - \frac{\partial \Sigma^\prime({\bf k},\omega)}{\partial \omega}\bigg\vert_{\omega = 0}, 
\end{equation}
where $\Sigma({\bf k},\omega)=\Sigma^\prime({\bf k},\omega) + \mathrm{i}\Sigma^{\prime\prime}({\bf k},\omega)$ is the complex-valued self-energy on the real frequency axis and $m({\bf k})$ is the bare electron mass. We obtain the derivative of the self-energy on the real axis using the Matsubara self-energy using the expression \cite{Arsenault2012benchmark}
\begin{equation}\label{eq:SEderivative}
    \lim_{\omega_n \rightarrow 0}\frac{\Sigma^{\prime\prime}({\bf k},\omega_n)}{\omega_n}
    = 
    \frac{\partial \Sigma^\prime({\bf k},\omega)}{\partial \omega}\bigg\vert_{\omega = 0}. 
\end{equation}
Eq.~\eqref{eq:SEderivative} can be easily proven using the Kramers-Kronig relations and is exact in the $\beta \rightarrow \infty$ limit. Here we approximate the left-hand side of Eq.~\eqref{eq:SEderivative} with its value at the lowest Matsubara frequency such that 
$m^*({\bf k})/m \approx 1 - \beta \Sigma^{\prime\prime}({\bf k},\omega_n=\pi/\beta)/\pi$. Finally, to obtain the effective mass value on the Fermi surface, we interpolate the Green's function data as described in Appendix~\ref{sec:effective_mass}.  

\begin{figure}[t]
    \centering
    \includegraphics[width=\columnwidth]{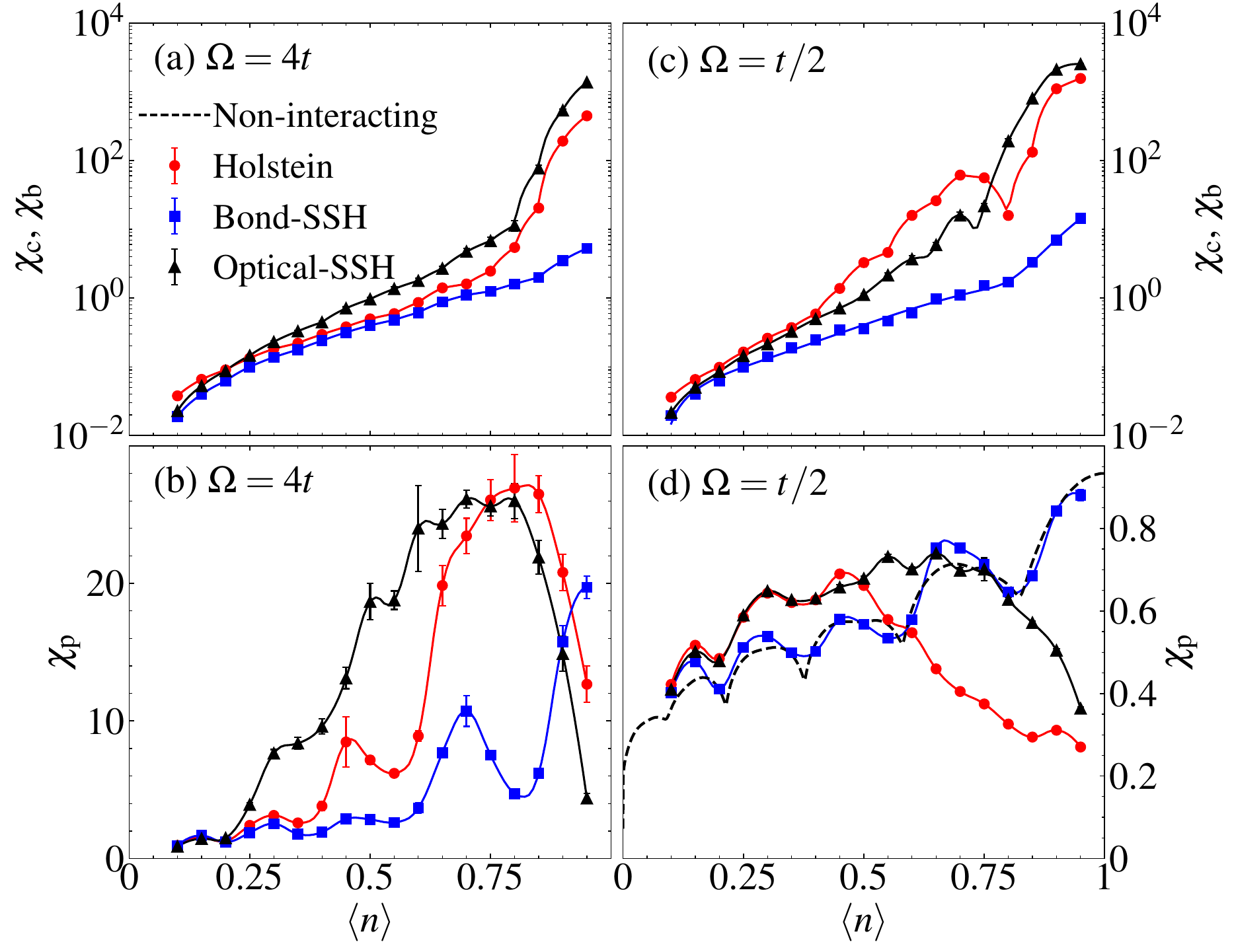}
    \caption{DQMC results for pairing and dominant charge/bond correlations obtained on $14\times 14$ lattices at $\beta = 16/t$ and $\lambda = 0.3$. 
    Panels (a) \& (c) show results for the CDW $[\chi_\mathrm{c}(\pi,\pi)]$ and BOW $[\chi_\mathrm{b}(\pi,\pi)]$ susceptibilities. Panels (b) and (d) show singlet $s$-wave pair-field $[\chi_\mathrm{p}({\bf q} = 0)]$ susceptibilities for the Holstein (red $\bigcirc$), and bond- (blue $\square$), and optical-SSH (black $\triangle$) models. Results for $\Omega = 4t$ ($t/2$) are shown in the left (right) column. The dashed black line in panel (d) is the pair-field susceptibility for the non-interacting model. Note that the charge and bond susceptibilities are plotted on a logarithmic scale, while the pair-field susceptibility is plotted on a linear scale.   
    \label{fig:density_sweeps}}
\end{figure}

\section{Results}  
Figure~\ref{fig:density_sweeps} compares the $s$-wave pair-field $\chi_\mathrm{p}(\mathbf{q})$, charge $\chi_\mathrm{c}(\mathbf{q})$, and bond-order $\chi_\mathrm{b}(\mathbf{q})$ susceptibilities for the three models as a function of filling $\langle n\rangle$. 
Results are shown for $\Omega = 4t$ [panels (a),(b)] and $t/2$ [panels (c),(d)] and were obtained on an $N=14\times 14$ cluster with fixed $\beta = 16/t$, and $\lambda = 0.3$. We have found that the uniform $s$-wave pairing correlations ${\bf q} = 0$ are the dominant pairing signal for all parameters examined here. In contrast, the dominant \gls*{CDW} and \gls*{BOW} correlations appear at the $\mathbf{q}=(\pi,\pi)$ ordering vector near half-filling. Therefore, we focus on these ordering vectors throughout this work. 

Focusing first on the charge and bond correlations, we find that the Holstein and optical-SSH models are dominated by \gls*{CDW} and \gls*{BOW} correlations for $0.75 \lesssim \langle n \rangle \le 1$, consistent with prior work~\cite{Nosarzewski2021superconductivity, Bradley2021superconductivity}. These orders compete directly with superconductivity in each case and become stronger as the phonon energy decreases. Conversely, the bond-SSH model is characterized by nearly degenerate antiferromagnetic (not shown), superconducting, and \gls*{CDW} correlations near half-filling in the $\Omega = 4t$ case, consistent with prior results for our chosen parameters~\cite{Feng2022phase, Xing2021quantum, Cai2022robustness, Gotz2022valence}. 

The \gls*{CDW}/\gls*{BOW} correlations in all three models are suppressed upon doping away from $\langle n \rangle = 1$ and eventually overtaken by the superconducting correlations. The strongest pairing correlations for the Holstein and optical-SSH models in the antiadiabatic limit ($\Omega = 4t$, Fig.~\ref{fig:density_sweeps}b) occur around $\langle n \rangle \approx 0.7 - 0.75$, and have broad dome-like dependence on the carrier concentration. In this case, the suppression of pairing near half-filling is due to competition with the \gls*{BOW}/\gls*{CDW} correlations, while the suppression for small $\langle n\rangle$ is due to the decreased density of carriers. This dome-like behavior is reminiscent of the bismuthates~\cite{Sleight2015bismuthates} and other quantum materials, where superconductivity is found near a competing order. Notably, the pairing correlations for the optical-SSH model are comparable to the Holstein model at intermediate doping but decay slower as the band is depleted. 

The pairing correlations in the bond-SSH model are notably smaller and exhibit slightly different dependence on the filling. The largest pairing correlations for this model occur at $\langle n \rangle = 1$. There is also a second narrow peak in $\chi_\mathrm{p}({\bf q} = 0)$ centered at $\langle n \rangle \approx 0.7$, where the pairing correlations are larger than the \gls*{BOW} correlations. (The exact position of this peak depends on the cluster size, see Appendix \ref{sec:finite_size}.) 

The superconducting correlations decrease dramatically when the phonon energy is lowered to $\Omega = t/2$ (Fig.~\ref{fig:density_sweeps}d), which is naturally expected for a phonon-mediated pairing mechanism where $T_\mathrm{c}\propto \Omega$. At this temperature ($\beta = 16/t$), the pairing correlations of the bond-SSH model are nearly identical to the non-interacting pair-field susceptibility (indicated by the dashed black line) for all $\langle n \rangle$. Conversely, the correlations in the Holstein and optical-SSH models are slightly enhanced over the non-interacting values at lower carrier concentrations $\langle n \rangle \lesssim 0.5$; however, $\chi_\mathrm{p}({\bf q} = 0)$ in both models is similar, suggesting that these models will have comparable $T_\mathrm{c}$'s at these carrier concentrations. Moving towards half-filling, $\chi_\mathrm{p}$ is suppressed below the non-interacting values for both the Holstein and optical-\gls*{SSH} models, but the optical model's pairing correlations persist to comparatively higher carrier concentrations. 

\begin{figure}[t]
    \centering
    \includegraphics[width=\columnwidth]{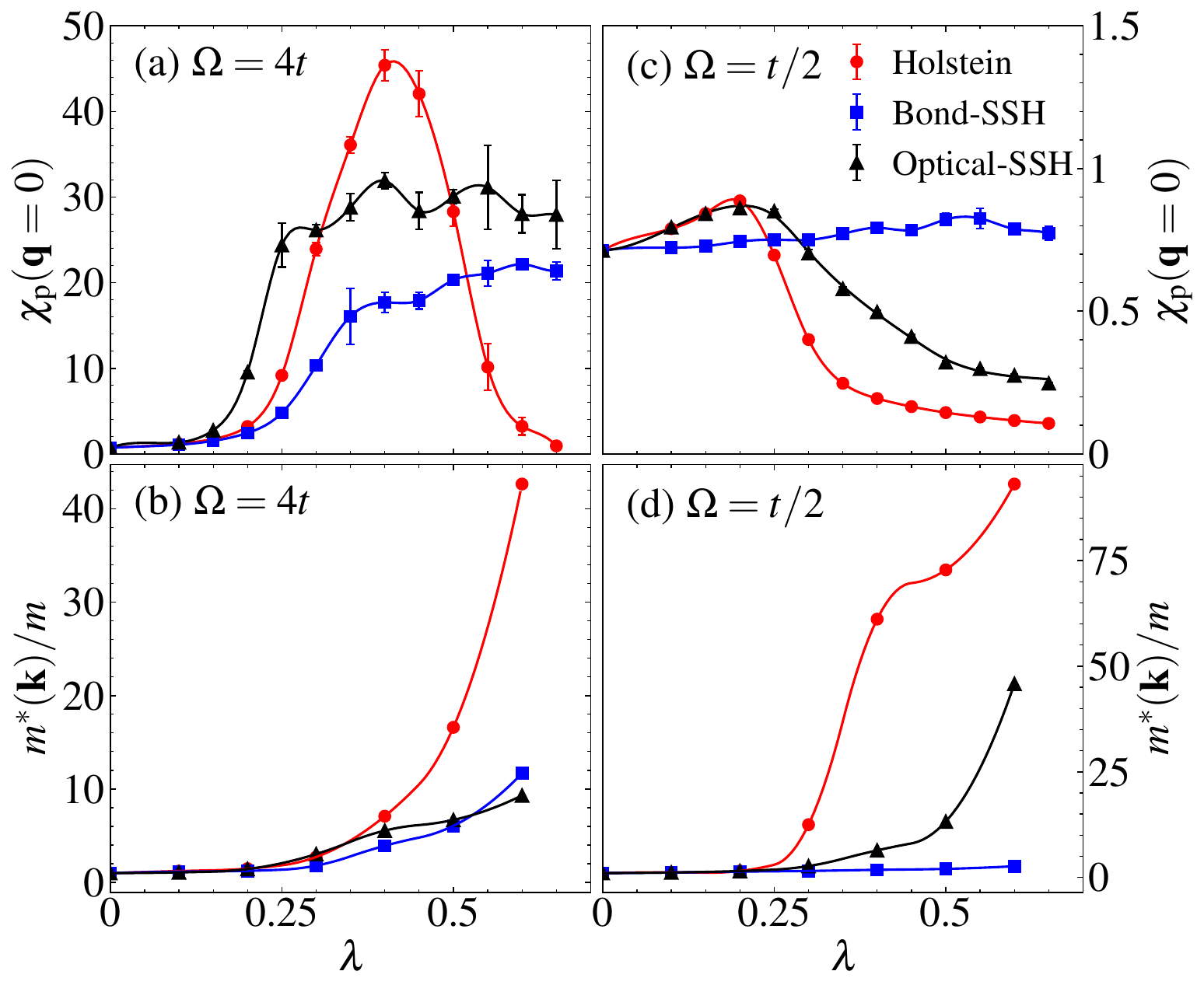}
    \caption{DQMC results for the Holstein (red $\bigcirc$), and bond-, (blue $\square$), and optical- (black $\triangle$) SSH models. The top row plots the $s$-wave pair-field susceptibility $\chi_\mathrm{p}({\bf q} = 0)$ as a function of $\lambda$ for (a) $\Omega = 4t$ and (c) $\Omega = t/2$. Panels (b) and (d) plot estimates of the effective mass $m^*({\bf k})/m$ measured on the Fermi surface at ${\bf k} = (k_\mathrm{F},0)$. All results were obtained from $N = 14\times 14$ lattices with $\beta = 16/t$ and $\langle n\rangle = 0.7$. 
    }
    \label{fig:lambda_sweeps}
\end{figure}

Figure~\ref{fig:lambda_sweeps}a explores how the superconducting correlations depend on $\lambda$. Here we fix $\beta = 16/t$, $\Omega=4t$, and $\langle n\rangle = 0.7$, where we observed strong pairing tendencies across all three models without a corresponding competing order. The growth of the pairing correlations in the Holstein model is non-monatomic, with a maximum value occurring at $\lambda \approx 0.4$ before being rapidly suppressed at large $\lambda$ by the formation of heavy bipolarons~\cite{Bradley2021superconductivity, Nosarzewski2021superconductivity}. This interpretation is supported by the behavior of the effective mass shown in Fig.~\ref{fig:lambda_sweeps}b. Here, we plot the effective mass at the Fermi momentum along the $(0,0)-(0,\pi)$ cut through the first Brillouin zone, where we observe the strongest band renormalizations (see also Appendix~\ref{sec:effective_mass}). In this case, the sharp drop in the pairing correlations coincides with a rapid increase in $m^*({\bf k}_\mathrm{F})/m$. 

Turning to the \gls*{SSH} models, we find that the pairing correlations exhibit a milder dependence on $\lambda$ in the strong coupling limit; in both the bond and optical models, $\chi_\mathrm{p}(\mathbf{q}=0)$ initially grows rapidly at $\lambda \le 0.2-0.3$ before leveling off to a constant value at larger couplings. At the same time, the effective mass for both models increases at a much slower rate in comparison to the Holstein model. This behavior demonstrates that the  \gls*{SSH} (bi)polarons remain much lighter to stronger values of $\lambda$~\cite{Sous2018light}, even for the finite carrier concentration considered here. 

Figs.~\ref{fig:lambda_sweeps}(c),(d) shows analogous results for $\Omega = t/2$. In this case, the pairing correlations at weak $\lambda$ for all three models are comparable to the non-interacting values. As $\lambda$ increases, $\chi_\mathrm{p}(\mathbf{q}=0)$ for the Holstein and optical-\gls*{SSH} models increase slightly before being rapidly suppressed with a concomitant rise in the effective mass, as in the antiadiabatic case. In contrast, the pairing correlations in the bond-SSH model are not suppressed at larger values of the $e$-ph coupling but are only slightly enhanced around $\lambda \approx 0.5$. The effective mass of the bond model remains small at all coupling values.

\begin{figure}[t]
    \centering
    \includegraphics[width=0.6\columnwidth]{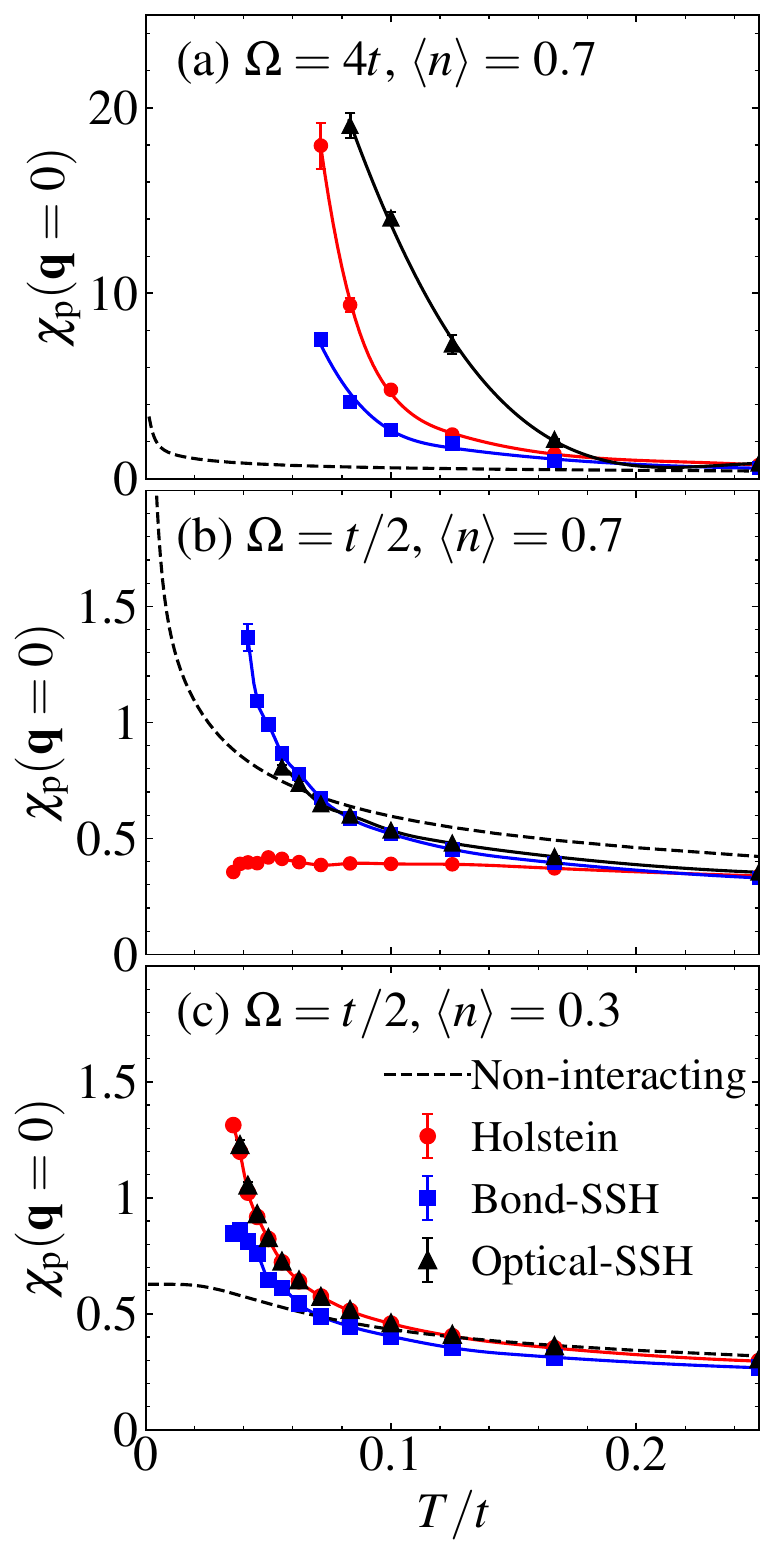}
    \caption{A comparison of the temperature dependence of the $s$-wave pair-field susceptibilities for the Holstein (red $\bigcirc$), and bond- (blue $\square$), and optical-SSH (black $\triangle$) models at fixed $\lambda = 0.3$ and $\langle n \rangle = 0.7$. Results are shown for (a) $\Omega = 4t$, $\langle n\rangle=0.7$, (b) $\Omega = t/2$, $\langle n\rangle=0.7$, and (c) $\Omega = t/2$, $\langle n\rangle=0.3$. All results were obtained on an $N = 14\times 14$ cluster. The solid lines are spline fits to the data and act as a guide to the eye. The dashed black lines are the pair-field susceptibility for the non-interacting model. \label{fig:temp_sweeps}}
\end{figure}

The strength of the pairing correlations at fixed temperature can be a poor proxy for the superconducting $T_\mathrm{c}$ that is ultimately realized in a system~\cite{Mai2021pairing}. For example, the growth of an initially subdominant pairing channel can sometimes overtake a dominant one as the system is cooled. For this reason, Fig.~\ref{fig:temp_sweeps} compares $\chi_\mathrm{p}(\mathbf{q}=0)$ as a function of $T$ for the three models, again at a fixed $\lambda = 0.3$. For $\Omega = 4t$ and $\langle n \rangle = 0.7$ (Fig.~\ref{fig:temp_sweeps}a), the optical-SSH model has the largest pairing correlations while the bond-SSH model has the smallest. However, the growth of $\chi_\mathrm{p}(\mathbf{q}=0)$ for the Holstein model appears to be outpacing the optical-SSH model. Taken at face value, this would suggest that the Holstein model has the highest $T_\mathrm{c}$ at this filling. However, finite size effects may slow the growth of the superconducting correlations in the optical model if the correlation length has become comparable to the cluster size. Regardless, the results in Fig.~\ref{fig:temp_sweeps}a indicate that the optical-SSH and Holstein models have comparable values of $T_\mathrm{c}$ for these parameters.

Fig.~\ref{fig:temp_sweeps}b shows results for $\Omega = t/2$ and $\langle n \rangle = 0.7$. Here, the pairing correlations for the Holstein model are significantly reduced as a function of temperature by the competing \gls*{CDW} correlations and local bipolaron formation~\cite{Esterlis2018breakdown, Nosarzewski2021superconductivity, Bradley2021superconductivity}. Conversely, the pairing correlations in the optical- and bond-SSH models continue to grow as the temperature is lowered and ultimately become larger than the non-interacting values. Both SSH models have comparable $\chi_\mathrm{p}(T)$ values as a function of temperature, implying they will have a similar $T_\mathrm{c}$. 

The behavior reported in Fig.~\ref{fig:temp_sweeps}b demonstrates that the SSH interactions are less prone to forming heavy bipolarons and can mediate superconductivity more effectively than the Holstein model in proximity to its competing CDW/BOW order. However, it is unclear if the transition temperature ultimately realized by the SSH models is larger than what could be achieved in the Holstein model if one could somehow suppress the competing \gls*{CDW} correlations. To address this question, Fig.~\ref{fig:temp_sweeps}c now examines $\chi_\mathrm{p}({\bf q} = 0)$ versus $T$ for the three models, this time focusing on $\Omega = t/2$ and $\langle n \rangle = 0.3$, far from any competing \gls*{CDW} correlations. In this case, the pairing correlations in all three models grow above the non-interacting value as the temperature is lowered. However, the strength of the pairing correlations for the Holstein and optical-SSH models are identical (within error bars), while the correlations in the bond-SSH model are smaller. We can thus reasonably conclude that the superconducting $T_\mathrm{c}$ for the optical-SSH and Holstein models are comparable for these parameters, while $T_\mathrm{c}$ for the bond-SSH model is smaller.

\begin{figure}[t]
    \centering
    \includegraphics[width=\columnwidth]{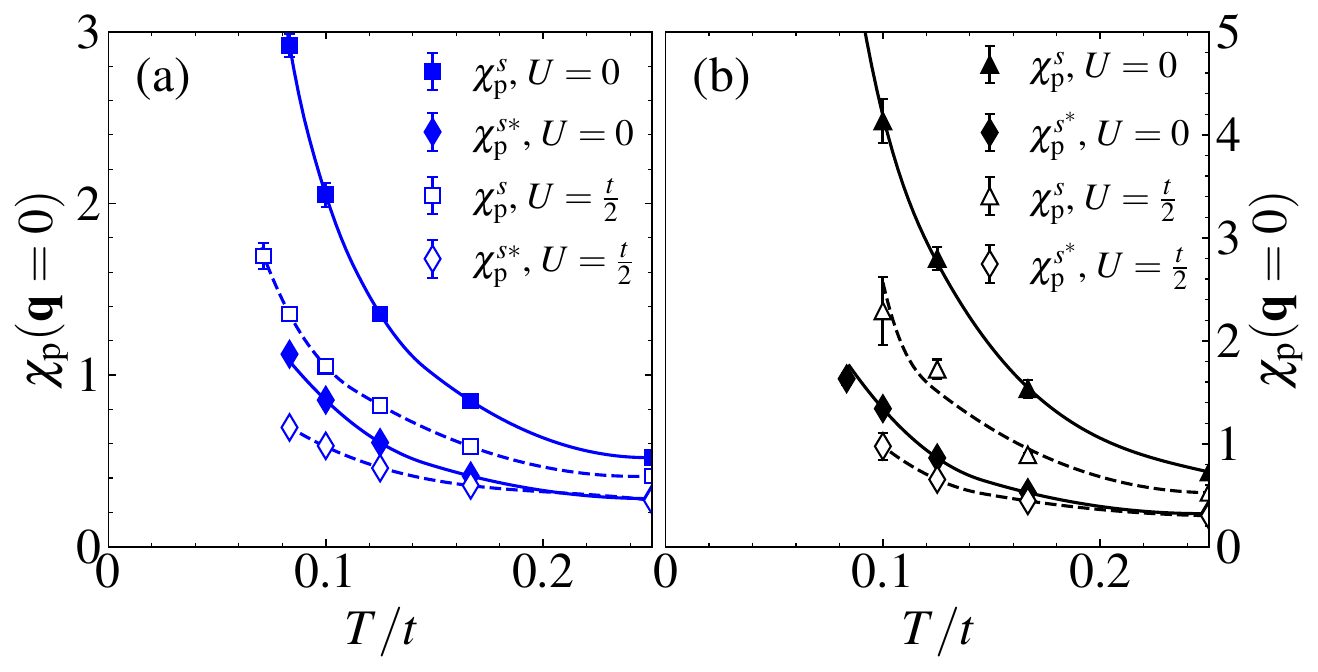}
    \caption{The temperature dependence of the uniform $s$-wave [$\chi_{\rm p}^{s}({\bf q} = 0)$, $\square$ points] and extended $s$-wave [$\chi_{\rm p}^{s^*}({\bf q} = 0)$, $\diamond$ points] pair-field susceptibilities for the (a) bond- and (b) optical-SSH models with a Hubbard $U = 0$ and $t/2$. All results were obtained on an $N = 8\times 8$ cluster due to the Fermion sign problem (see Appendix~\ref{sec:fermion_sign}) and fixed $\langle n\rangle = 0.5$, $\Omega = 4t$, and $\lambda = 0.3$.\label{fig:Hubbard}}
\end{figure}

Finally, we turn to the question of how robust the pairing correlations are against the inclusion of the $e$-$e$ interaction. To this end, Fig.~\ref{fig:Hubbard} plots the superconducting correlations in the bond- and optical-Hubbard-SSH models, where we have added a slight on-site Hubbard repulsion of $U =t/2$. Since the non-zero value of $U$ induces a Fermion sign problem (see Appendix~\ref{sec:fermion_sign}), here we focus on results obtained on a smaller $N=8\times 8$ lattice with $\Omega = 4t$, $\lambda = 0.3$, and $\langle n\rangle = 0.5$, which corresponds to the filling with the strongest pairing correlations on this size lattice. For both the bond (Fig.~\ref{fig:Hubbard}a) and optical-SSH (Fig.~\ref{fig:Hubbard}b) models, we find that the pairing correlations decrease rapidly once a small Hubbard repulsion is included in the model. 

\section{Discussion}
We have used \gls*{DQMC} to study and contrast the ordering tendencies of the \gls*{2D} Holstein and bond-/optical-SSH models over a wide range of carrier concentrations. Our results reveal that these models produce different qualitative behaviors depending on the filling of the underlying band. Close to half-filling, the Holstein and optical \gls*{SSH} models are dominated by \gls*{CDW} and \gls*{BOW} order, respectively. The superconducting correlations were generally suppressed near these competing orders; however, the suppression was more severe for the Holstein model. Conversely, the bond-SSH model supported superconducting correlations up to half-filling, especially in the anti-adiabatic limit $(\Omega = 4t)$. 
Variational calculations in the two-particle limit~\cite{Sous2018light} have shown that the SSH (bi)polarons are lighter and more mobile than the Holstein bipolarons for a larger range of couplings. Our results demonstrate that this property persists to finite carrier concentrations and allows the \gls*{SSH} models to support superconductivity in regions of phase space where the polaronic effects prevent pairing for the Holstein model. This result lends support to the hypothesis that \gls*{SSH} interactions can circumvent some factors that limit pairing in the conventional Holstein and Fr{\"o}hlich models, at least in proximity to the competing \gls*{BOW} order. 

At more dilute concentrations, where competition with \gls*{CDW} or \gls*{BOW} correlations is absent, we find that the pairing correlations for the optical and bond-SSH models are comparable to that of the Holstein model. This result suggests that the bond- and optical-\gls*{SSH} models will not produce significantly larger superconducting transition temperatures than that of the Holstein model, at least for carrier concentrations larger than $\langle n \rangle \ge 0.3$. 

Another critical issue in studying \gls*{SSH}-like interactions is the potential role of the sign changes in the effective hopping integrals. The models considered here and elsewhere treat the nonlinear dependence of the hopping integrals on the bond distance using the linear approximation. This treatment allows the sign of the effective hopping integral to change as the atoms vibrate about their equilibrium positions, which can dramatically affect the ground and excited states of the model~\cite{Nocera2021bipolaron, banerjee2023ground}. Indeed, we have found that these sign changes frequently occur in our \gls*{DQMC} simulations when the dimensionless coupling $\lambda$ or phonon energy $\Omega$ is large (see Appendix~\ref{sec:sign_change}). We have allowed these changes to occur in the current work since we were interested in studying the ordering tendencies in the linear SSH models as they are commonly formulated. However, future studies should explore how additional nonlinear $e$-ph interactions or anharmonic lattice potentials might alter the \gls*{BOW} and superconducting correlations observed here. Such terms have already been shown to substantially alter the physics of the Holstein model~\cite{Li2015effects, Dee2020relative, Sous2021phonon, Paleari2021quantum, kovac2023optical}.

\section*{Acknowledgements} 
We thank M. Berciu and R. T. Scalettar for insightful discussions and comments on this paper. This work was supported by the U.S. Department of Energy, Office of Science, Office of Basic Energy Sciences, under Award Number DE-SC0022311. It used resources of the Oak Ridge Leadership Computing Facility, which is a DOE Office of Science User Facility supported under Contract No. DE-AC05-00OR22725.

\appendix
\section{Dimensionless coupling parameter}\label{sec:lambda}

The dimensionless coupling for a momentum-dependent $e$-ph interaction $g_\nu({\mathbf k},{\mathbf q})$ is defined  \cite{Allen1983theory} as 
\begin{equation}\label{eq:lambda}
    \lambda = 2\mathcal{N}(0)\sum_\nu\left\llangle \frac{|g_\nu(\mathbf{k},\mathbf{q})|^2}{\Omega_{\nu,\mathbf{q}}}\right\rrangle_\mathrm{FS}, 
\end{equation}
where $\nu$ is a mode index, $\xi_{\bf k} = \epsilon_{\bf k}-\mu$ with $\epsilon_\mathbf{k} = -2t\sum_\nu \cos(k_\nu a)$ is the non-interacting electron dispersion measured relative to the chemical potential, $\mathcal{N}(0) = \tfrac{1}{N}\sum_\mathbf{k} \delta(\xi_{\bf k})$ is the density of states at the Fermi level per spin species, and $\llangle \cdot \rrangle_\mathrm{FS}$ denotes a \gls*{FS} average.

In the 2D Holstein model, the electrons couple to a single dispersionless phonon branch $\Omega_{\bf q} = \Omega$ via a momentum-independent $e$-ph coupling $g_\mathrm{h} = \alpha_\mathrm{h}/\sqrt{2M_\mathrm{h}\Omega}$. In this case, the dimensionless coupling reduces to $\lambda_\mathrm{h} = \frac{\alpha_\mathrm{h}^2}{M_\mathrm{h}\Omega^2 W}$, where we have approximated $\mathcal{N}(0) \approx W^{-1}$, where $W = 8t$ the non-interacting bandwidth. 

In the 2D optical- and bond-SSH models, the electrons couple to two dispersionless optical phonon branches $\nu = x,y$ with 
$\Omega_{x,{\bf q}} = \Omega_{y,{\bf q}} = \Omega$ via momentum-dependent $e$-ph coupling constants  \cite{Li2011perturbation, Zhang2021Peierls} 
\begin{equation}\label{eq:gkq}
    \begin{aligned}
    g^{\rm o}_\nu(\mathbf{k},\mathbf{q}) &= 4 \frac{\alpha_{\rm o}}{\sqrt{2M_\mathrm{o}\Omega}}{\rm i} \sin\left(\tfrac{q_\nu a}{2}\right)\cos((k_\nu+q_\nu/2)a), \\
    g^{\rm b}_\nu(\mathbf{k},\mathbf{q}) &= 2 \frac{\alpha_{\rm b}}{\sqrt{2M_\mathrm{b}\Omega}} e^{{\rm i}q_\nu a/2} \cos((k_\nu+q_\nu/2)a).
    \end{aligned}
\end{equation}
for the optical- and bond-SSH models, respectively, where we have taken $\hbar = 1$. The dimensionless coupling is obtained by performing the \gls*{FS} average as defined in Eq.~\eqref{eq:lambda}. However, due to the finite size of our clusters, performing accurate \gls*{FS} averages is difficult, particularly away from half-filling. Therefore, we approximate Eq.~\eqref{eq:lambda} by replacing the \gls*{FS} average with a simple average over the first Brillouin zone. The corresponding dimensionless $e$-ph couplings are $\lambda_\mathrm{b} = \frac{4\alpha_\mathrm{b}^2}{M_\mathrm{b}\Omega^2 W}$ and $\lambda_{\rm o} = \frac{8\alpha_{\rm o}^2}{M_{\rm o}\Omega^2 W}$ for the bond- and optical-SSH models, respectively. To ensure that the value of the dimensionless coupling remains the same among the three models, we fix the ratios of the microscopic couplings to $\alpha^2_\mathrm{o} = \alpha^2_\mathrm{H}/8$ and $\alpha^2_\mathrm{b} = \alpha^2_\mathrm{H}/4$. Note that our definition for $\lambda_\mathrm{b}$ differs from Ref.~\cite{Zhang2023bipolaronic} by a factor of two. Therefore, the momentum dependence of $g_\nu^\mathrm{b}({\bf k},{\bf q})$ was neglected and the ratio $\alpha^2_\mathrm{b} = \alpha^2_\mathrm{H}/2$ was adopted. For this reason, the strength of the coupling to the bond model is stronger in their analysis. 

\section{Sign changes in the effective hopping integral}\label{sec:sign_change}
\begin{figure}[t]
    \centering
    \includegraphics[width=\columnwidth]{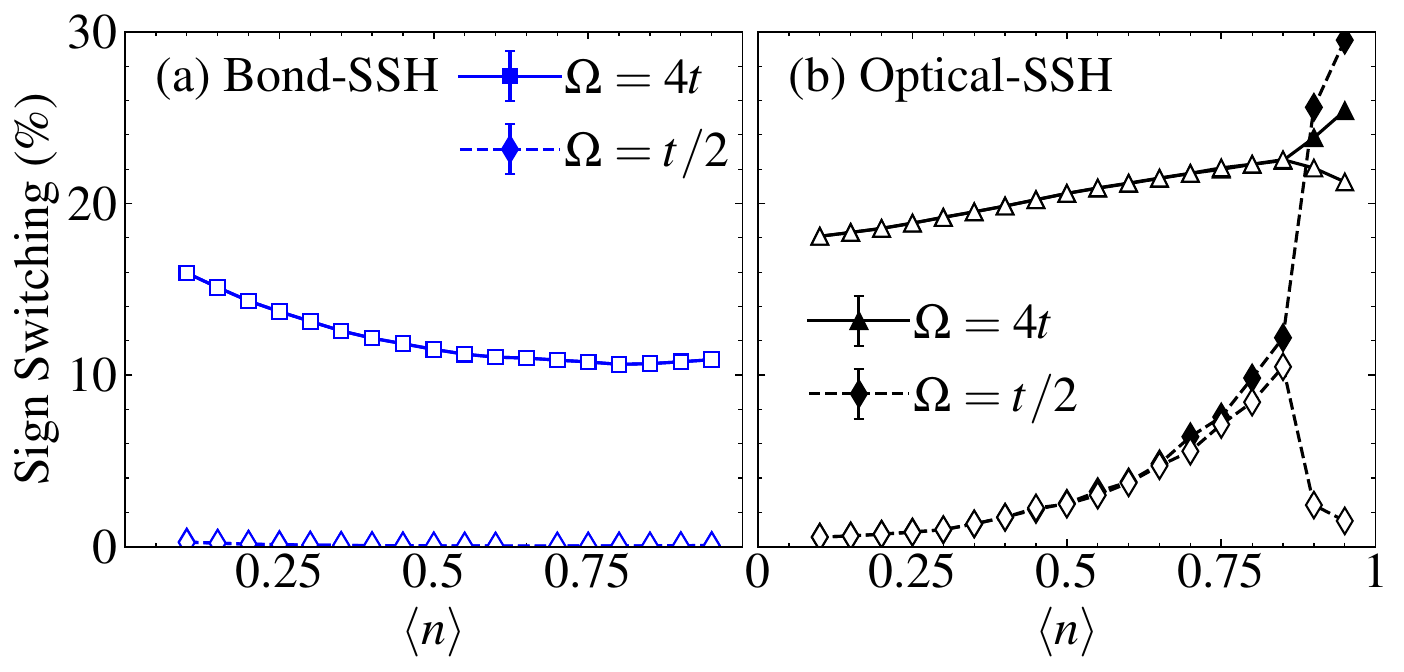}
    \caption{Sign switching of the hopping integral as a function of filling for (a) the bond-SSH model in the x-direction (closed $\square$) and in the y-direction (open $\square$). Also shown are complementary results for (b) the optical-SSH model (closed and open $\triangle$). All results for $\lambda = 0.3$ at $\beta t = 16$ on an $N = 14\times 14$ cluster.
    \label{fig:lattice_sgn_switch}}
\end{figure}

Both the optical- and bond-\gls*{SSH} models approximate the nonlinear dependence of the hopping integrals on the atomic displacements within the linear approximation. At this level, both models allow the sign of the effective hopping integral to change when the lattice displacements become sufficiently large, which is frequently overlooked in non-perturbative numerical simulations. In principle, one can suppress the frequency of these sign changes by including nonlinear $e$-ph interactions~\cite{Adolphs2013going} or additional anharmonic terms in the lattice potential~\cite{Paleari2021quantum}, which tend to reduce the size of the lattice displacements found in the linear model. However, these terms can significantly affect the pairing and charge correlations in the linear model~\cite{Li2015effects} and prevent comparisons to works that didn't include these interactions. For this reason, we have chosen to simulate the optical- and bond-\gls*{SSH} models as they are formulated in the linear approximation and not artificially prevented Monte Carlo updates that produce a sign change in the effective hopping. 

To assess the severity of this problem, Fig.~\ref{fig:lattice_sgn_switch} plots the percentage of times the effective hopping changes sign during our \gls*{DQMC} simulations for both the bond- and optical-\gls*{SSH} models. Note that we do \textit{not} average the phonon positions over imaginary time before calculating this percentage. Results are shown here for $\Omega = 4t$, $\beta = 16/t$, $N = 14\times 14$, and as a function of $\langle n\rangle$. 
Sign switching occurs more frequently in the optical model for fixed $\lambda = 0.3$, which can be understood once one recognizes that the displacement of two atoms controls the bond distance in the optical model. The bifurcation of the results for the optical-\gls*{SSH} model is due to forming the BOW phase, where inequivalent bond lengths appear along the $x$- and $y$- directions. \\

\section{Effective mass estimates}\label{sec:effective_mass}
\begin{figure}[t]
    \centering
    \includegraphics[width=0.8\columnwidth]{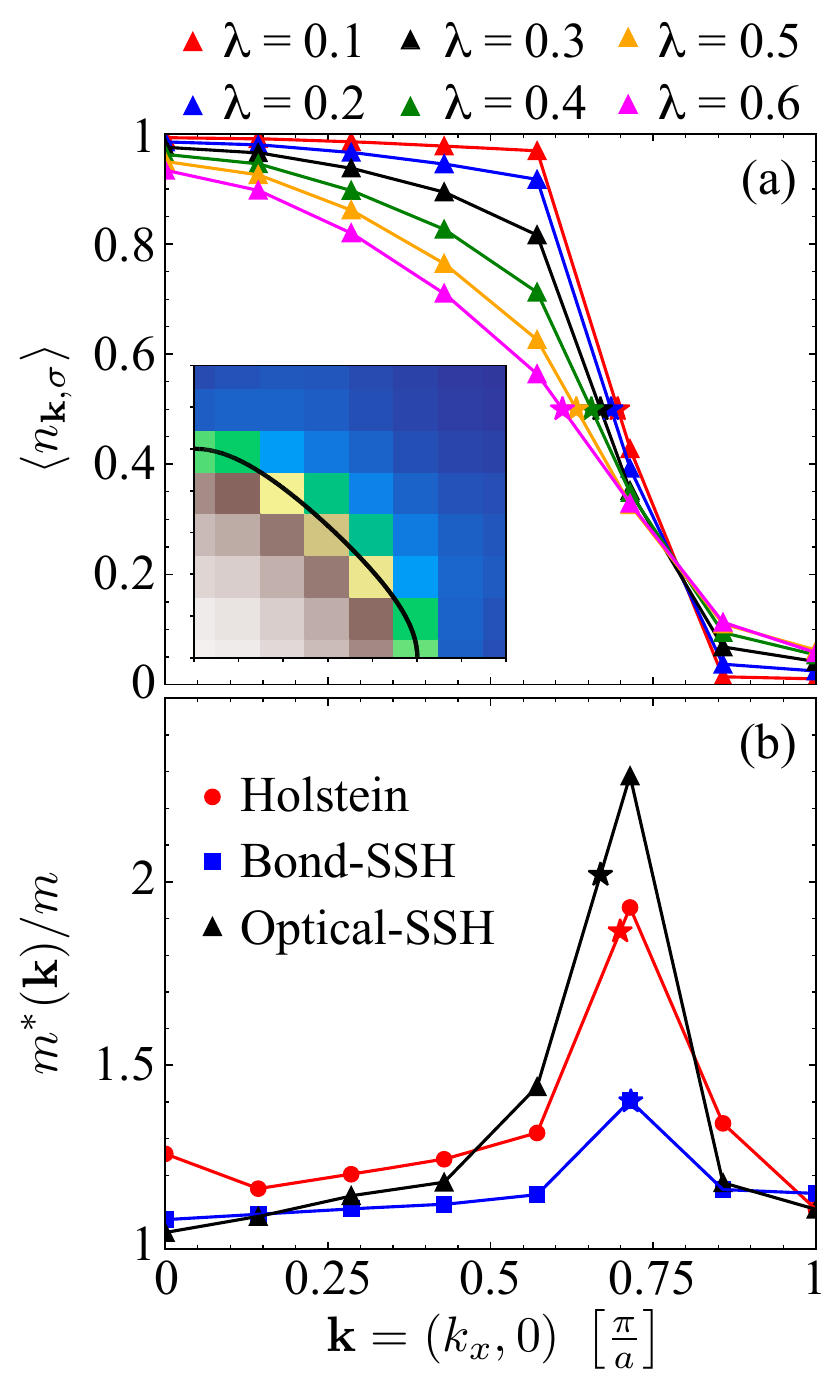}
    \caption{\label{fig:effective_mass} 
    (a) The expectation value of the number operator $\langle n_{{\bf k},\sigma}\rangle$ for the optical-SSH model as a function of $\lambda$, plotted along the ${\bf k} = (k_x,0)$ direction of the first Brillouin zone (FBZ). The allowed momentum points in the cluster are indicated by  the triangles. The Fermi momentum, indicated by the stars, is determined by $\langle n_{{\bf k},\sigma}\rangle = 1/2$ and estimated by linearly interpolating the DQMC data. 
    The inset shows $\langle n_{{\bf k},\sigma}\rangle$ in the upper quadrant of the FBZ for $\lambda = 0.3$, where the solid black line indicates the location of the non-interacting Fermi surface. 
    (b) The effective mass $m^*({\bf k})/m$ along $(k_x,0)$ for the Holstein (red $\bigcirc$), optical- (blue $\square$), and bond-SSH (black $\triangle$) models for $\lambda = 0.3$. 
    The value of $m^*({\bf k}_\mathrm{F})/m$ (indicated by the stars) is estimated using linear interpolation. 
    All data were obtained on $N = 14\times 14$ clusters with $\beta = 16/t$ and $\Omega = 4t$.
    }
\end{figure}

As outlined in Sec.~\ref{sec:methods}, the effective mass $m^*({\bf k})$ is obtained from our DQMC data 
using the relationship
\begin{equation*}
    \frac{m^*({\bf k})}{m} \approx 1 - \frac{\beta \Sigma^{\prime\prime}({\bf k},\omega_n)}{\pi},  
\end{equation*}
where $m$ is the electron's bare mass in the non-interacting limit and $\Sigma({\bf k},\omega_n) = \Sigma^\prime({\bf k},\omega_n) + \mathrm{i}\Sigma^{\prime\prime}({\bf k},\omega_n)$ is the complex-valued self-energy on the Matsubara frequency axis. ($m$ also depends on ${\bf k}$ for our tight-binding model, so $\frac{m^*({\bf k})}{m}$ should be understood including the ${\bf k}$-dependence of both $m^*$ and $m$.) 
To obtain the self-energy, we measure the unequal imaginary time Green's function $G_\sigma({\bf k},\tau)$ which is then Fourier transformed to the Matsubara frequency axis $G_\sigma({\bf k},\omega_m)$ using the discrete Lehmann representation~\cite{Kaye2022discrete}. To accomplish this task, we use the  \texttt{Lehmann.jl} package of the \texttt{libdlr} library~\cite{Kaye2022libdlr}. The self-energy is then obtained by inverting Dyson's equation $G^{-1}({\bf k},\omega_n) = {\rm i}\omega_n - \xi_{\bf k} - \Sigma({\bf k},\omega_n)$.  

We extract the value of the effective mass on the Fermi surface by linearly interpolating our DQMC data, as illustrated in Fig.~\ref{fig:effective_mass}. First, we determine the location of the Fermi surface using the condition $\langle n_{{\bf k}_\mathrm{F},\sigma}\rangle = 1-G_\sigma({\bf k}_\mathrm{F},0) = 0.5$, where $G_\sigma({\bf k},0)$ is the equal-time Green's function (Fig~\ref{fig:effective_mass}a). We then interpolate the $\frac{m^*({\bf k})}{m}$ data to obtain its value on the Fermi surface (Fig~\ref{fig:effective_mass}b). 

\section{The Fermion sign problem}\label{sec:fermion_sign}
\begin{figure}[t]
    \centering
    \includegraphics[width=0.7\columnwidth]{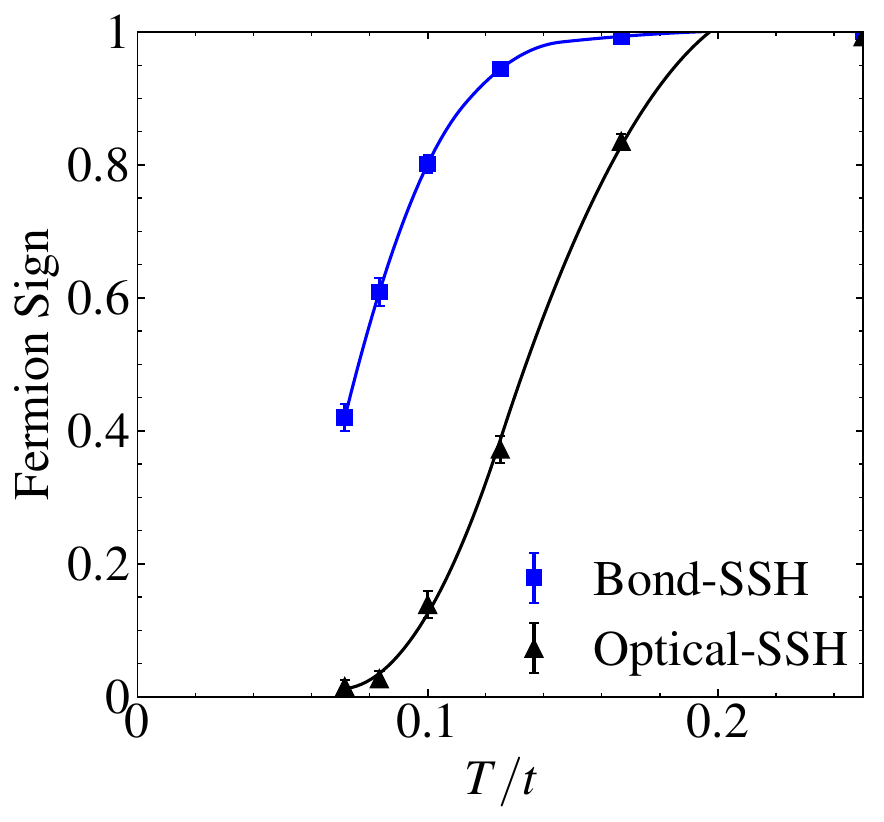}
    \caption{   
    \label{fig:fermion_sign} The average value of the Fermion sign obtained for the simulations with $U = t/2$ 
    presented in Fig.~\ref{fig:Hubbard} of the main text. }
\end{figure}

The Holstein and \gls*{SSH} models considered in this work can be simulated with \gls*{DQMC} without a sign problem. However, all three models develop a sign problem once we include the Hubbard interaction~\cite{Johnston2013determinant}. (Only the optical-SSH model at half-filling remains sign-problem-free.) For reference, Figure~\ref{fig:fermion_sign} plots the average value of the Fermion sign obtained for the simulations presented in Fig.~\ref{fig:Hubbard} with a finite $U = t/2$. Interestingly, the bond-\gls*{SSH} model appears to have a less severe sign problem than the optical-\gls*{SSH} model. 

\section{Finite size effects}\label{sec:finite_size}
\begin{figure}[t]
    \centering
    \includegraphics[width=\columnwidth]{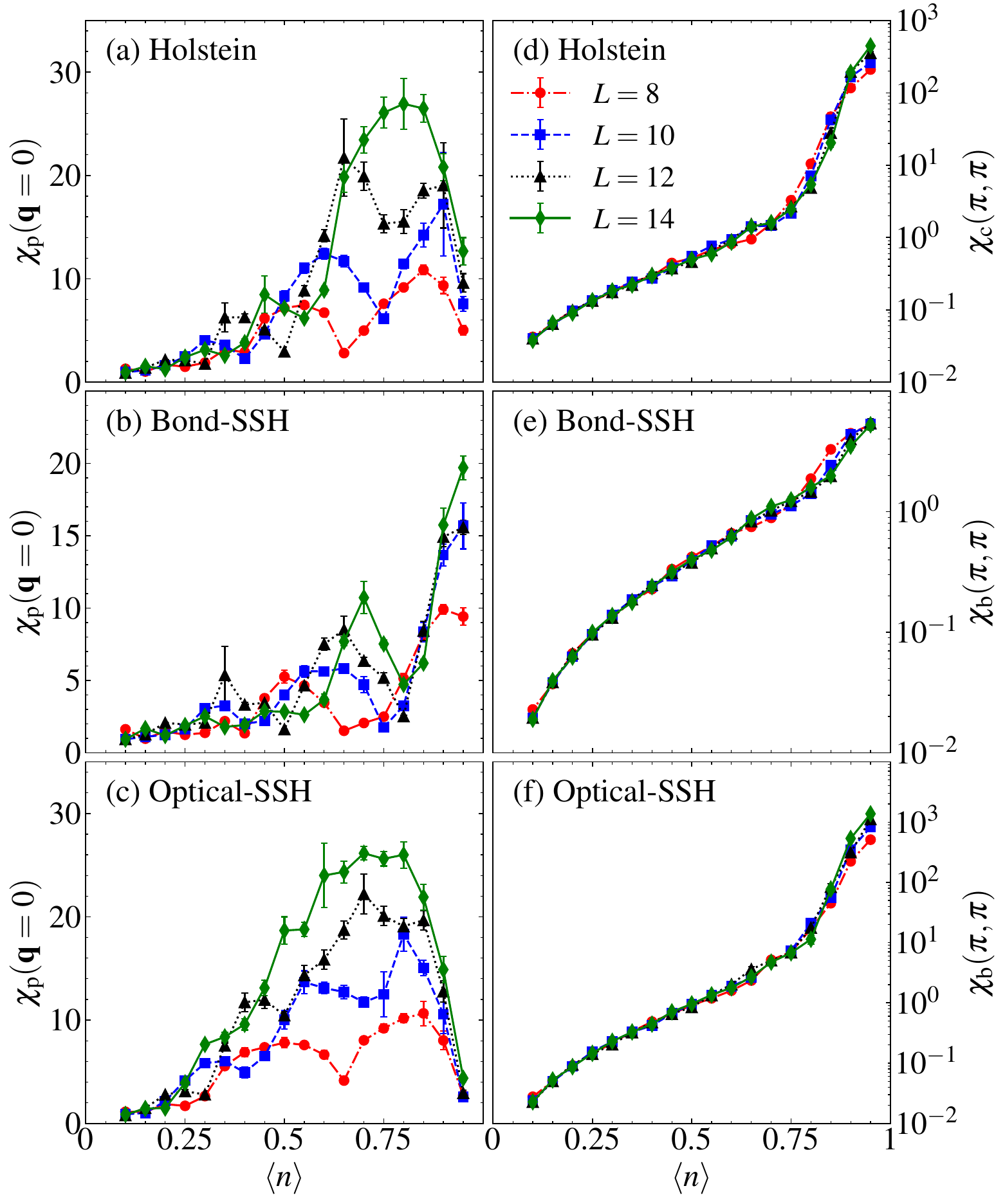}
    \caption{   
    \label{fig:finite_size} All panels are DQMC results for $\Omega = 4t$, $\lambda = 0.3$ at $\beta t = 16$ on different $L\times L$ lattices. Panels (a) and (b) show the uniform (${\bf q} = 0$) $s$-wave pair-field $\chi_{\rm p}(\mathbf{q})$ and $\mathbf{q}=(\pi,\pi)$ charge-density-wave $\chi_{\rm c}({\bf q})$ susceptibilities for the Holstein model. Panels (c) and (d) show $\chi_{\rm p}$ and $\mathbf{q}=(\pi,\pi)$ bond-ordered-wave $\chi_{\rm b}(\mathbf{q})$ susceptibility for the bond-SSH model. Similarly, panels (e) and (f) show $\chi_{\rm p}$ and $\chi_{\rm b}$ for the optical-SSH model. }
\end{figure}

Figure~\ref{fig:finite_size} assesses the finite size effects for our simulations of the Holstein, bond-, and optical-SSH models. Here we fix the simulation parameters to $\Omega = 4t$, $\lambda = 0.3$, and $\beta = 16/t$ and vary carrier concentrations over $0 < \langle n \rangle < 1$. (Our model is particle-hole symmetric because we only include the nearest-neighbor hopping $t$.) The left column of Fig.~\ref{fig:finite_size} plots the uniform $s$-wave pair-field susceptibility for the Holstein [Fig.~\ref{fig:finite_size}a], bond-SSH [Fig.~\ref{fig:finite_size}b], and optical-SSH [Fig.~\ref{fig:finite_size}c] models, respectively. Similarly, the right column of Fig.~\ref{fig:finite_size} plots each model's dominant charge/bond order. Specifically, Fig.~\ref{fig:finite_size}d plots the ${\bf q} = (\pi,\pi)$ charge correlations for the Holstein model, while Figs.~Fig.~\ref{fig:finite_size}e and Fig.~\ref{fig:finite_size}f plot the ${\bf q} = (\pi,\pi)$ bond susceptibilities for the bond- and optical-SSH models, respectively. The results indicate that all three models exhibit noticeable finite-size effects; however, the overall qualitative behavior remains the same across all system sizes. As the cluster size increases, there is an overall increase in the strength of the pairing correlations, which is accompanied by a shift of the peak position toward half-filling. At the same time, the strength of the \gls*{CDW} (\gls*{BOW}) correlations in the Holstein (optical-SSH) model near $\langle n \rangle = 1$ increase with cluster size, consistent with the presence of long-range order at this filling. Conversely, the strength of the \gls*{BOW} correlations in the bond-SSH model exhibits a much weaker finite size dependence, consistent with short-range correlations. Finally, all three models' charge and bond susceptibilities show a weaker dependence on cluster size for fillings less than $\langle n \rangle \lesssim 0.75$. 

\bibliography{references}
\end{document}


\title{Supplementary materials for ``A comparative study of the pairing correlations in the Holstein and optical Su–Schrieffer–Heeger models''}
\date{\today}

\author{Andy Tanjaroon Ly}
\affiliation{Department of Physics and Astronomy, The University of Tennessee, Knoxville, TN 37996, USA}
\affiliation{Institute of Advanced Materials and Manufacturing, The University of Tennessee, Knoxville, TN 37996, USA\looseness=-1} 

\author{Sohan Malkaruge Costa}
\affiliation{Department of Physics and Astronomy, The University of Tennessee, Knoxville, TN 37996, USA}
\affiliation{Institute of Advanced Materials and Manufacturing, The University of Tennessee, Knoxville, TN 37996, USA\looseness=-1} 

\author{Benjamin~Cohen-Stead}
\affiliation{Department of Physics and Astronomy, The University of Tennessee, Knoxville, TN 37996, USA}
\affiliation{Institute of Advanced Materials and Manufacturing, The University of Tennessee, Knoxville, TN 37996, USA\looseness=-1} 

\author{Richard~T.~Scalettar}
\affiliation{Department of Physics, University of California, Davis, California 95616, USA}

\author{Steven Johnston}
\affiliation{Department of Physics and Astronomy, The University of Tennessee, Knoxville, TN 37996, USA}
\affiliation{Institute of Advanced Materials and Manufacturing, The University of Tennessee, Knoxville, TN 37996, USA\looseness=-1}

{
\let\clearpage\relax
\maketitle
}
\section*{Supplementary Note 1: Dimensionless electron-phonon coupling constant}

For a momentum dependent electron-phonon ($e$-ph) coupling constant that depends on the electron ($\mathbf{k}$) and phonon ($\mathbf{q}$) momenta, the general parameterization of the dimensionless coupling constant is
\begin{equation}
    \lambda = 2\mathcal{N}(0)\left\llangle \frac{g(\mathbf{k},\mathbf{q})}{\Omega(\mathbf{q})} \right\rrangle_{\mathrm{FS}}
\end{equation}
where $\mathcal{N}(0)$ is the density of states at the Fermi level and $\llangle\cdot\rrangle_{\mathrm{FS}}$ denotes a Fermi surface (FS) average. For the Holstein model, this constant is typically parameterized as $\lambda_\mathrm{h} = 2g^2/\Omega W = \alpha^2/\Omega^2 W$ where $g(\mathbf{k},\mathbf{q})\approx g = \alpha/\sqrt{2M\Omega}$ and $\mathcal{N}(0)\approx 1/W = 1/8t$ is the non-interacting bandwidth.

For the bond-/optical-SSH models, we approximate the FS average by instead performing a simple average over the Brillouin zone:

\begin{subequations}
\begin{align}
  \lambda^{\mathrm{BZ}}_{\mathrm{b}} & = \frac{2\alpha^2}{\Omega^2 W} \\
  \lambda^{\mathrm{BZ}}_{\mathrm{o}} & = \frac{4\alpha^2}{\Omega^2 W}.
\end{align}
\end{subequations}

\newpage
\section*{Supplementary Note 2: Finite size effects}

\begin{figure}[h]
    \centering
    \includegraphics[width=0.85\textwidth]{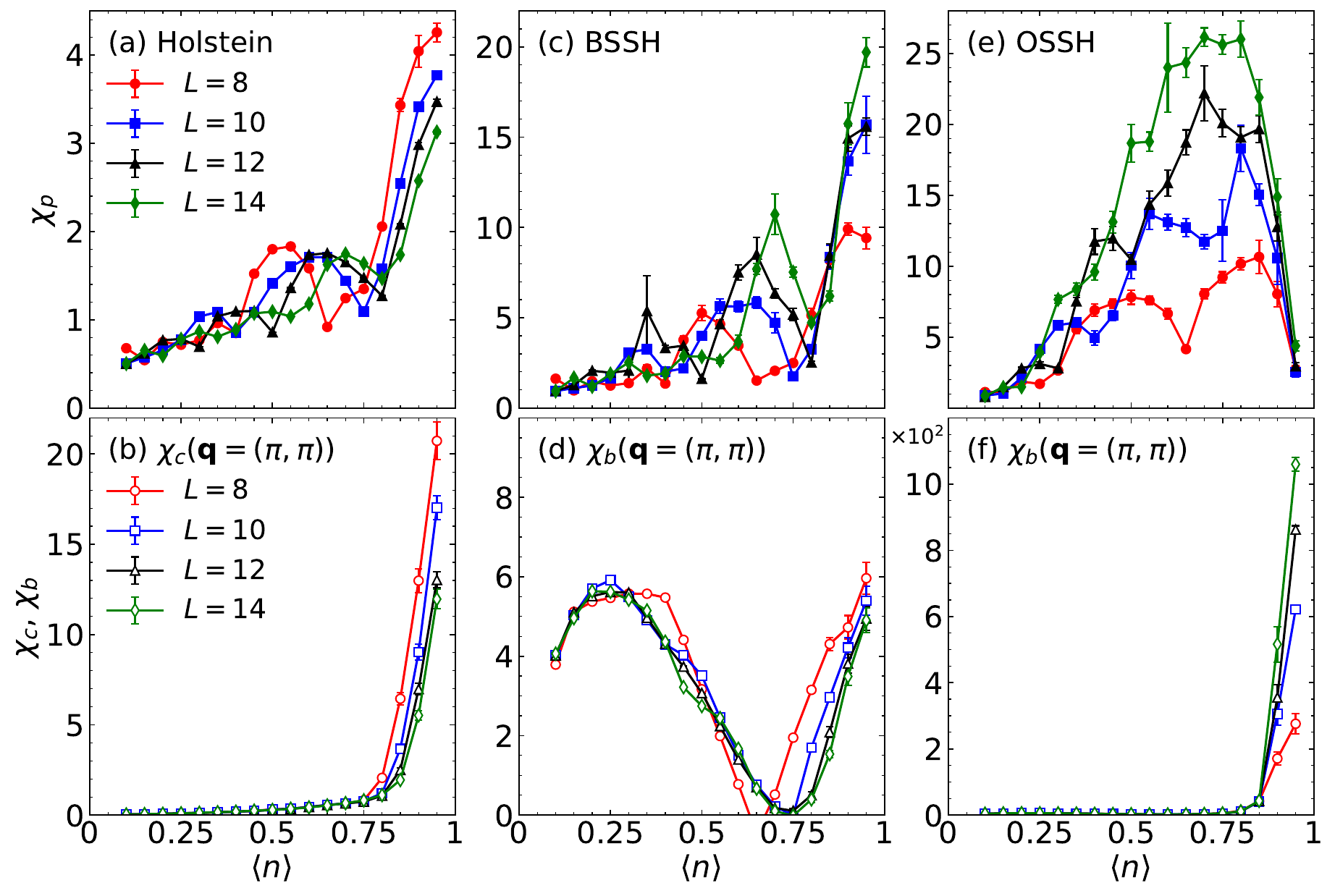}
    \caption{\textcolor{red}{Needs to be replaced.}
    All panels are DQMC results for $\Omega = 4t$, $\lambda = 0.15$ at $\beta t = 16$ on different $L\times L$ lattices. Panels (a) and (b) show pair-field $\chi_p$ and charge density wave $\chi_{c}$ susceptibilities for the Holstein model. Panels (c) and (d) show bond ordered wave $\chi_{b}$ susceptibility and  $\chi_p$ for the bond SSH model. Similarly, panels (e) and (f) show $\chi_{b}$ and $\chi_p$ for the optical SSH model.   
    \label{fig:lattice_sweeps}}
\end{figure}

Figure~\ref{fig:lattice_sweeps} assesses the finite-size effects for simulations of the Holstein and bond/optical SSH models. Here we fix the simulation parameters to $\Omega/t = 4$, $\lambda = 0.15$, and $\beta t = 16$ and vary the electron concentration over $0 < \langle n \rangle < 1$. (Our model is particle-hole symmetric because we only include the nearest-neighbor hopping $t$.) The top row of Fig.~\ref{fig:lattice_sweeps} plots the uniform $s$-wave pair-field susceptibility for the Holstein [Fig.~\ref{fig:lattice_sweeps}(a)], bond-SSH [Fig.~\ref{fig:lattice_sweeps}(c)], and optical-SSH [Fig.~\ref{fig:lattice_sweeps}(e)] models, respectively. Similarly, the bottom row of Fig.~\ref{fig:lattice_sweeps} plots the dominant competing charge/bond order in the respective models. Specifically, Fig.~\ref{fig:lattice_sweeps}(b) plots the ${\bf q} = (\pi,\pi)$ charge correlations for the Holstein model, while Figs.~Fig.~\ref{fig:lattice_sweeps}(d) and Fig.~\ref{fig:lattice_sweeps}(f) plot the ${\bf q} = (\pi,\pi)$ bond susceptibilities for the bond- and optical-SSH models respectively.  

Figure~\ref{fig:lattice_sweeps} indicates that while all three models exhibit finite size effects, the qualitative behaviors are consistent across all cluster sizes. 

\newpage 
\section*{Supplementary Note 3: SSH sign switching}
\begin{figure}[h]
    \centering
    \includegraphics[width=0.8\textwidth]{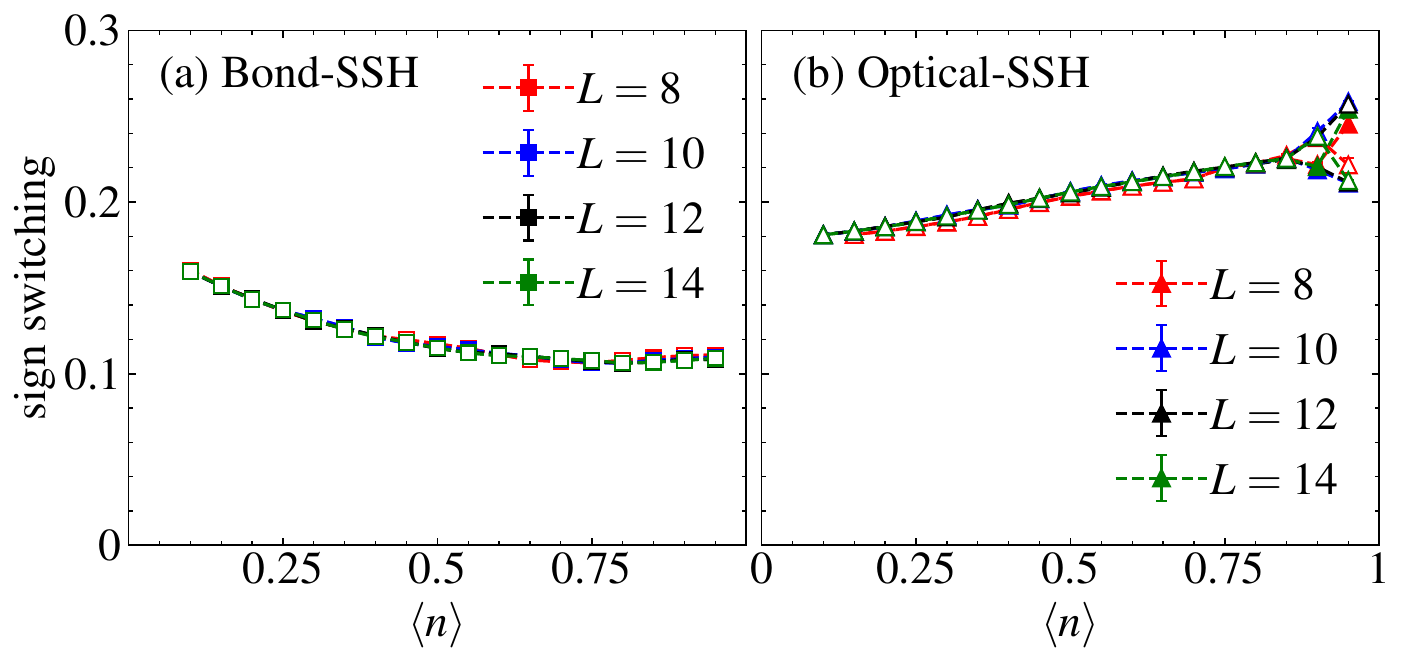}
    \caption{All panels are DQMC results for $\Omega = 4t$, $\lambda = 0.3$ at $\beta t = 16$ on different $L\times L$ lattices as a function of filling. Panel (a) shows the sign switching of the hopping integral for the bond SSH model in the x-direction (closed $\square$) and in the y-direction (open $\square$). Similarly, panel (b) shows the sign switching for the optical SSH model in the x-direction (closed $\triangle$) and in the y-direction (open $\triangle$).   
    \label{fig:lattice_sgn_switch}}
\end{figure}

\newpage 
\section*{Supplementary Note 4: The Fermion sign problem}
\begin{figure}[h]
    \centering
    \includegraphics[width=0.45\textwidth]{./Figures/FigureS3.pdf}
    \caption{   
    \label{fig:fermion_sign}}
\end{figure}

\clearpage
\bibliography{references}